\documentclass[review,12pt]{elsarticle}
\usepackage{lineno,hyperref}
\modulolinenumbers[5]

\journal{Journal of \LaTeX\ Templates}

\usepackage{amsmath}
\usepackage{graphicx}

\title{Nonlinear Conductivity in Graphene }

\author[rvt]{S. S. Abukari\corref{cor1}}
\author[rvt]{S. Y. Mensah}
\author[els]{R. Musah}
\address[rvt]{Department of Physics, College of Agriculture and Natural Sciences, U.C.C, Ghana.}
\cortext[cor1]{Corresponding author\\ Email: sulemana70@gmail.com\\
This work is licensed under the Creative Commons Attribution International\\ License (CC BY).
\url{http://creativecommons.org/licenses/by/4.0/} }
\author[focal]{ \\ N. G. Mensah}
\author[rvt]{ K. A. Dompreh}
\address[focal]{Department of Mathematics, College of Agriculture and Natural Sciences, U.C.C, Ghana}
\address[els]{Department of Applied Physics, University for Development Studies,
Navorongo, Ghana}

\begin{document}
\begin{abstract}
We consider the tight-binding approximation for the description of energy bands 
of graphene, together with the standard Boltzmann's transport equation and constant 
relaxation time, an expression for the conductivity was obtained. We predicted 
strong nonlinear effects in graphene which may be useful for high frequency 
generation.
\end{abstract}

\maketitle

\section{Introduction}
Graphene was discovered by Novoselov et al in 2004~\cite{1}-~\cite{3} and 
has since attracted a great deal of interest due to its unique properties 
like mechanical, electrical, thermal, etc arising from its highly symmetric 
two-dimensional honeycomb-lattice structure~\cite{4,5}. This makes graphene 
potentially applicable in carbon-based nanoelectronics`\cite{6, 7}.  The 
nonquadratic energy spectrum of graphene allows it to exhibit nonlinear 
phenomena~\cite{8} and its non-addivity property provides the mutual dependence 
of electron or hole motion in orthogonal directions~\cite{9}. Several unique 
physical properties of graphene have been studied both theoretically and 
experimentally~\cite{8}-~\cite{12}.\\

Investigation into electronic properties of graphene has revealed that the 
electron dispersion law is linear in momemtum near the Fermi points and 
thereby causing the quasiparticles to behave like massless Dirac Fermions~\cite{12}. 
However, the tendency of graphene to absorb impurities on its surface and 
interacts with the impurities electrons results in the formation of nonparabolicity 
of the energy band which makes the electronic properties of graphene essentially 
nonlinear under moderate electric fields~\cite{13}. This nonlinearity makes 
graphene to exhibit plethora of transport phenomena~\cite{12}-~\cite{20}. 
Under different conditions of an external electric field, an electron in 
graphene is predicted to reveal a variety of physical effects such as Bloch 
oscillations, self-induced transparency, absolute negative conductance, etc.
Electronic properties and electronic transport in graphene is the subject of 
many theoretical papers~\cite{12}-~\cite{20}.  Nevertheless, the electrodynamic 
properties of graphene is worth further studying because it is the basis for 
developing carbon-based devices. Using the kinetic transport equation, we shall 
in this work study the effect of high frequency (hf) conductivity in graphene 
by following the approaches of~\cite{21}-~\cite{26}. 

\section{Theory}
Proceeding as in references~\cite{21}-~\cite{26}, we consider the motion of an 
electron in the presence of high frequency electric field $E(t)$. The electric 
field $E(t)$  is directed along the graphene axis and the conductivity is 
derived using the Boltzmann kinetic equations describing electron transport 
in graphene for the distribution functions in the relaxation time approximation 
as follows:
\begin{eqnarray}
\frac{\partial f_{s}}{\partial t}+eE(t)\frac{\partial f_{a}}{\partial\mathbf{p}}=\frac{F-f_{s}}{\tau}
\end{eqnarray}
\begin{eqnarray}
\frac{\partial f_{a}}{\partial t}+eE(t)\frac{\partial f_{s}}{\partial\mathbf{p}}=\frac{-f_{a}}{\tau}
\end{eqnarray}
where $e$ is the electron charge, $\mathbf{p}$ is the electron dynamical momentum, $F$ is the equilibrium distribution function and $f_s$ as well as $f_a$ the symmetric and antisymmetric distribution functions respectively.
Solving Eqs.($1$) and ($2$) in a constant electric field yields,
\begin{eqnarray}
\frac{\partial^{2} f_{a}^{o}}{\partial\xi^{2}}-\chi_{o}^{2}f_{a}^{o}=\frac{\partial F}{(eEa\tau)\partial\xi}
\end{eqnarray}
where $\xi_{o}^{2}=\frac{\pi^{2}}{(eEa\tau)^{2}}$ and $\mathbf{p}=\frac{\pi}{a}\xi$.See~\cite{26}\\
Using $\frac{\partial F}{\partial\xi}=\frac{\partial F}{\partial\varepsilon}\frac{\partial\varepsilon}{\partial\xi}$,
 the solution to Eq.($2$) for the necessary boundary conditions $f_{a}(-1)=f_{a}(+1)=0$ 
is
\begin{eqnarray}
f^{o}_{a}=\varepsilon\frac{\partial F}{\partial\varepsilon}
\frac{eaE\tau}{(eaE\tau)^{2}+1}
\end{eqnarray}
The spectrum of electrons in graphene is given by [24,25].
\begin{eqnarray}
|\varepsilon(p)|=\pm\frac{3\gamma_{o}b}{2\hbar}|\mathbf{p}-\mathbf{p_{F}}|
\end{eqnarray}

where $\gamma_{0}\approx2.7eV$,$b=0.142nm$ is the distance between the neighbouring carbon atoms in the graphene. $+$ and $-$ signs are related to the conduction and valence bands respectively. With $p_F$ as the constant quasimomentum corresponding to the particular Fermi point.
The current density of the mobile electron in the $1^{st}$ Bz for graphene is given as [24, 25]
\begin{eqnarray}
j_{x}=\frac{2e}{(2\pi\hbar)^{2}}\int\int v_{x}f_{a}^{o}d^{2}\mathbf{p}
\end{eqnarray}
and the quasiclassical velocity $v_{x}(p)$  of an electron moving along the graphene axis can be
\begin{eqnarray}
v_{x}(p)=\frac{\partial\varepsilon}{\partial p}=\frac{a}{\pi}
\frac{\partial\varepsilon}{\partial\xi}=\frac{a}{\pi}grad\varepsilon 
\end{eqnarray}
and writing
\begin{eqnarray}
d^{2}\mathbf{p}=\frac{\pi}{a}d^{2}\xi=\frac{\pi}{a}\frac{ds}{|\frac{a}{\pi}grad\varepsilon|}dE
\end{eqnarray}
$ds$ is the element of the length of the curve.\\
Substituting Eqs.($3$),($6$) and($7$) into($5$) we get,
\begin{eqnarray}
j_{x}=-\frac{e^{2}}{2\pi(\hbar)^{2}}\frac{E\tau}{(eaE\tau)^{2}+1}
\int|\varepsilon|\frac{\partial F}{\partial\varepsilon}d\varepsilon
\end{eqnarray}
Using (4), we obtain
\begin{eqnarray}
j_x=\frac{8ln2}{\pi(\hbar)^{2}}\frac{\tau e^{2}k_{B}TE}{(\omega_{B}\tau)^{2}+1}
\end{eqnarray}
and
\begin{eqnarray}
\sigma_{x}=\frac{8ln2}{\pi(\hbar)^{2}}\frac{\tau e^{2}k_{B}T}{(\omega_{B}\tau)^{2}+1}
\end{eqnarray}
where $\omega_{B}\tau=eaE\tau$.  See!\cite{26}\\
By linearizing  Eqs. ($1$) for the perturbation
\begin{eqnarray}
E(t)=(E+E_{\omega,k}e^{-i(\omega t+kx)}),f_s=f_{s}^{o}+f_{s}^{1}
e^{-i(\omega t+kx)},f_a=f_{a}^{1}e^{-i(\omega t+kx)}
\end{eqnarray}
we obtain
\begin{eqnarray}
\frac{\partial^{2}f_{a}^{1}}{\partial\xi^{2}}-\chi_{\omega}^{2}f_{a}^{1}=\pi^{2}\frac{(i\omega\tau-1)}{(eaE\tau)^{2}}f_{a}^{o}\frac{E_{\omega k}}{E}-\frac{\partial^{2}f_{a}^{o}}{\partial\xi^{2}}\frac{E_{\omega k}}{E}
\end{eqnarray}
where
\begin{eqnarray}
\chi_{\omega}^{2}=\frac{\{1-2i\omega\tau-(\omega\tau)^{2}\}}{(eaE\tau)^{2}}
\end{eqnarray}
Using $\frac{\partial F}{\partial\xi}=\frac{\partial F}{\partial\varepsilon}\frac{\partial\varepsilon}{\partial\xi}$, the solution to Eq(11) for the necessary boundary conditions $f_{a}(-1)=f_{a}(+1)=0$ is
\begin{eqnarray}
f_{a}^{1}=\left[\varepsilon\frac{\partial F}{\partial\varepsilon}\left(\frac{(eaE\tau)}{(eaE\tau)^{2}+1}\right)\frac{E_{\omega k}}{E}\left\{\frac{i\omega\tau-1-(eaE\tau)^{2}}{(eaE\tau)^{2}-(\omega\tau)^{2}+1-2i\omega\tau}\right\}\right]
\end{eqnarray}
Substituting Eqs.($12$),($6$)and($7$) into($5$) and using ($4$), we obtain
we get,
\begin{eqnarray}
j_{x}=-\frac{e^{2}}{2\pi(\hbar)^{2}}\left(\frac{(eaE\tau)}{(eaE\tau)^{2}+1}\right)\left\{\frac{i\omega\tau-1-(eaE\tau)^{2}}{(eaE\tau)^{2}-(\omega\tau)^{2}+1-2i\omega\tau}\right\}\nonumber\\
\times\int|\varepsilon|\frac{\partial F}{\partial\varepsilon}d\varepsilon
\end{eqnarray}
where $\int\vert \varepsilon\vert\frac{\partial F}{\partial \varepsilon} d\varepsilon = K_{\beta}Tln2$
making Eqn.($16$) yield
\begin{eqnarray}
j_{x}=-\frac{K_{\beta}Tln2e^{2}}{2\pi(\hbar)^{2}}\left(\frac{(eaE\tau)}{(eaE\tau)^{2}+1}\right)\left\{\frac{i\omega\tau-1-(eaE\tau)^{2}}{(eaE\tau)^{2}-(\omega\tau)^{2}+1-2i\omega\tau}\right\}\nonumber\\
\end{eqnarray}
\begin{eqnarray}
\sigma_{x}(\omega)=\sigma_{o}\left\{\frac{i\omega\tau-1-(eaE\tau)^{2}}{(\omega_{B}\tau)^{2}-(\omega\tau)^{2}+1-2i\omega\tau}\right\}
\end{eqnarray}
where $eaE\tau=\omega_{B}\tau$ and $\sigma_{o}=\frac{2ln2\tau e^{2}k_{B}T}{\pi(\hbar)^{2}}\frac{\omega_{B}\tau}{(\omega_{B}\tau)^{2}+1}$. See [26]

\section{Results, Discussion and Conclusion}
We present the results of a kinetic equation approach of a graphene subject to constant electric field $E$. The electric field $E$  is directed along the graphene axis.  Exact expression for the direct conductivity was obtained in eq.(9). The nonlinearity is analyzed using the dependence of the normalized conductivity $j_x$ as a function of $E$. 
. Figure 1 illustrates the dependence of $j_x$ on $E$. The figure shows a linear dependence of $j_x$ on $E$ at weak values of $E$. As E increases, $j_x$  increases, and at a some value of $E$, $j_x$ reaches a maximum value. Further increase in $E$ results in the decrease of $j_x$. The slope of the curve is the dc differential conductivity, $\sigma_x=dj_{x}/dE$. See equation (10). If $E\approx1$, the Bloch frequency is in the terahertz frequency range and if $E>1$, $\sigma_x$ is negative and therefore graphene demonstrates negative differential conductivity (NDC).
Figure 2 elucidates the graph of conductivity component $(\sigma_{x}(\omega))/\sigma_{o}$ obtained in Eq. (14) on the normalized frequency $\omega/\omega_{B}$ . The dimensionless complex conductivity $(\sigma_x ω)/\sigma_o$ depends strongly and nonlinearly on the normalized frequency  $\sigma/\omega_B$ .  We observed that the real part of the complex conductivity will become more negative with increasing frequency, until a resonance minimum occurs just before the Bloch frequency $\omega_{B}\tau$. This negative-conductivity resonance close to the Bloch frequency makes the graphene an active medium for a Bloch oscillator without domain instabilities induced by negative dc conductivity. 
In summary, using the solution of the Boltzmann's transport equation with constant relaxation time $\tau$ approximation, we obtained an exact expression for the conductivity of graphene. We noted a strong nonlinear effets which may be useful for the generation of high frequemcy radiation.

\begin{figure}[h!]
	\centering
	\includegraphics[width = 7.5cm]{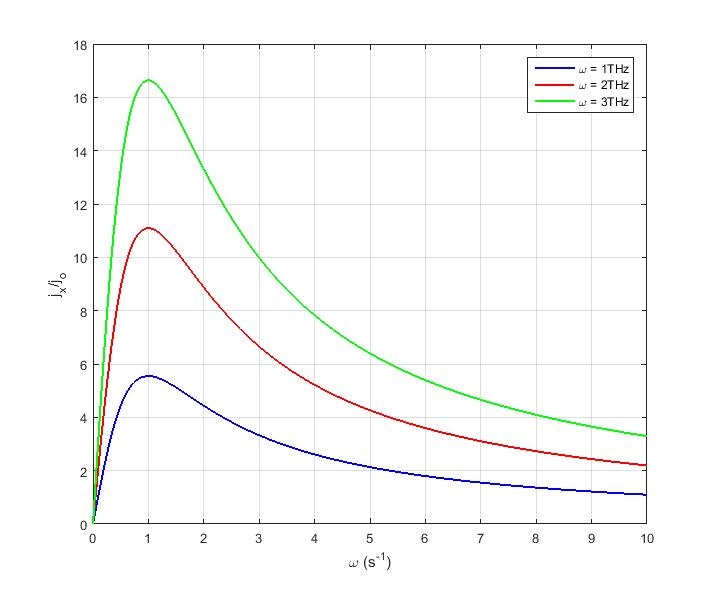}	    
 \caption{A plot of current density $j_x$ as a function of $E$  in graphene for expression ($9$).}
\end{figure}

\begin{figure}[h!]
	\centering
	\includegraphics[width = 7.0cm]{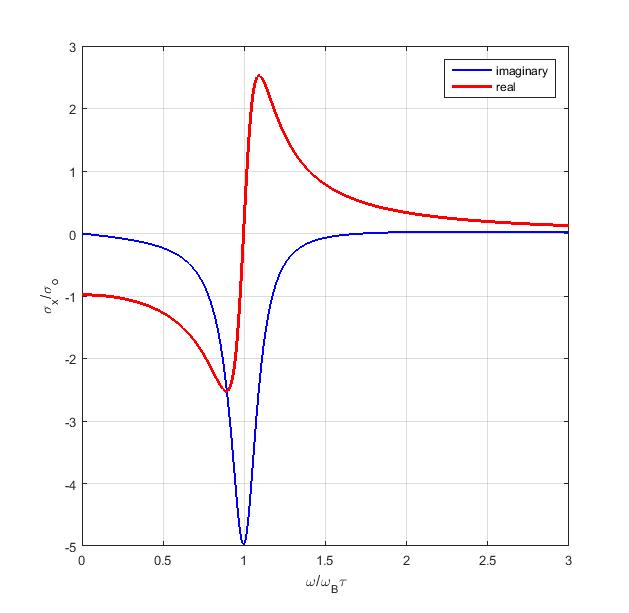}	    
 \caption{A plot of a normalized complex conductivity ($\sigma_x \omega)/\sigma_0$    as a 
function of dimensionless frequency $\omega/\omega_{\beta}$  in graphene for expression ($9$) 
when $\omega_{\beta}\tau =10$.}
\end{figure}

\end{document}